\documentclass[fleqn,usenatbib,useAMS]{mnras}

\usepackage{graphicx}	
\usepackage{amsmath}	
\usepackage{amssymb}	
\usepackage{multicol}        
\usepackage{bm}		
\usepackage{pdflscape}	

\usepackage{hyperref}
\usepackage{xspace}

\usepackage[T1]{fontenc}
\usepackage{ae,aecompl}
\usepackage{newtxtext,newtxmath}

\newcommand{\otherstar}{donor\xspace}

\title[SN Ia ejecta-\otherstar{} interaction]{Type Ia supernova ejecta-\otherstar{} interaction: explosion model comparison}

\author[McCutcheon, C. et al.]{
C. McCutcheon,$^{1,2,3}$
Y. Zeng,$^{1,2,4}$
Z.-W. Liu,$^{1,2}$\thanks{E-mail: zwliu@ynao.ac.cn}
R. G. Izzard,$^{3}$\thanks{E-mail: r.izzard@surrey.ac.uk}
K.-C. Pan,$^{5}$\thanks{E-mail: kuochuan.pan@gapp.nthu.edu.tw}
H.-L. Chen,$^{1,2}$
and Z. Han$^{1,2,4}$\\
\\
$^{1}$Yunnan Observatories, Chinese Academy of Sciences (CAS), Kunming 650216, P.R. China\\
$^{2}$Key Laboratory for the Structure and Evolution of Celestial Objects, CAS, Kunming 650216, P.R. China\\
$^{3}$Astrophysics Research Group, Faculty of Engineering and Physics, University of Surrey, Guildford, GU2 7XH, United Kingdom\\
$^{4}$University of Chinese Academy of Sciences, Beijing 100049, China\\
$^{5}$Department of Physics and Institute of Astronomy, National Tsing Hua University, Hsinchu 30013, Taiwan, R.O.C.\\
}

\date{Accepted XXX. Received YYY; in original form ZZZ}

\pubyear{2022}

\begin{document}

\label{firstpage}
\pagerange{\pageref{firstpage}--\pageref{lastpage}}
\maketitle

\begin{abstract}

In the single-degenerate scenario of Type Ia supernovae (SNe Ia), the interaction between high-speed ejected material and the \otherstar{} star in a binary system is expected to lead to mass being stripped from the donor. A series of multi-dimensional hydrodynamical simulations of ejecta-\otherstar{} interaction have been performed in previous studies most of which adopt either a simplified analytical model or the W7 model to represent a normal SN Ia explosion. Whether different explosion mechanisms can significantly affect the results of ejecta-\otherstar{} interaction is still unclear. In this work, we simulate hydrodynamical ejecta interactions with a main-sequence (MS) \otherstar{} star in two dimensions for two near-Chandrasekhar-mass explosion models of SNe Ia, the W7 and N100 models. We find that about $0.30$ and $0.37\,\mathrm{M}_{\sun}$ of hydrogen-rich material are stripped from a $2.5\,\mathrm{M}_\odot$ \otherstar{} star in a $2\,\mathrm{day}$ orbit by the SN Ia explosion in simulations with the W7 deflagration and N100 delayed-detonation explosion model, respectively. The \otherstar{} star receives a kick  of about $74$ and $86\,\mathrm{km}\,\mathrm{s}^{-1}$, respectively, in each case. The modal velocity, about $500\,\mathrm{km}\,\mathrm{s}^{-1}$, of stripped hydrogen-rich material in the N100 model is faster than the W7 model, {with modal velocity of about $350\,\mathrm{km\,s^{-1}}$}, by a factor~1.4. Based on our results, we conclude that the choice of near-Chandrasekhar-mass explosion model for normal SNe Ia seems to not significantly alter the ejecta-\otherstar{} interaction for a given main-sequence \otherstar{} model, at least in 2D.  \\

\end{abstract}



\section{Introduction}
\label{sec:introduction}

Type Ia supernovae (SNe~Ia) are widely believed to be thermonuclear explosions of carbon-oxygen white dwarfs (CO~WDs, \citealt{Hoyle1960}). As a class, they exhibit remarkable homogeneity in light-curves and are extremely luminous, making them ideal candidates to be used as standard candles for distance measurement on an universal scale. This property was notably used to first measure the accelerating expansion of the universe \citep{Riess1998, Schmidt1998,Perlmutter1999}. They are also important contributors to the abundance of heavy elements and therefore chemical evolution in their host galaxies, making the subject of their progenitor systems of great importance in the field. However, there has been much debate over the nature of their progenitor systems in trying to explain the variation in observed properties, leading to a range of potential progenitor models. For instance, the single-degenerate (SD) scenario  \citep{Whelan1973}, the double-degenerate (DD) scenario \citep{Iben1984, Webbink1984, Pakmor2010}  and the sub-Chandrasekhar double-detonation scenario  \citep{Taam1980, Fink2007, Shen2010, Sim2010, Woosely2011, Gronow2020} have been proposed for producing SNe~Ia. In addition, the exact explosion mechanism of SNe~Ia remains unclear, although they are accepted to be thermonuclear explosions of WDs in binary systems \citep{Hillebrandt2013,Maoz2014}.

The SD and the DD scenarios have been widely studied over the past decades. In the DD scenario, two WDs with total mass $\geq M_{\mathrm{Ch}}\sim1.4\,\mathrm{M}_{\sun}$ merge because of gravitational wave emission in a binary or possibly through direct collision in a multiple system  \citep{Iben1984, Webbink1984, Raskin2009,Rosswog2009}. This model finds explains the lack of observed hydrogen and helium lines in the late-time spectra of most SNe Ia and the delay-time distribution matches the observations well \citep{Maoz2014}. However, because the range of WD masses,  about $0.7-1.1\,\mathrm{M}_\odot$, which could merge to produce SNe Ia is large, the DD scenario struggles to explain the observed homogeneity of SN~Ia light-curves. In addition, it is unclear whether a slow WD-WD merger produces a type~Ia supernova. It is instead possible that an accretion-induced collapse forms a  neutron star \citep{Nomoto1985}.

In the SD scenario, a WD accretes material from a main-sequence (MS), sub-giant (SG), red-giant (RG) or helium (He) star in a close binary system until the WD mass is close to the Chandrasekhar mass limit,  $M_\mathrm{Ch}$, at which point the star explodes as an SN~Ia \citep{Whelan1973, Han2004,Liu2018,Liu2020aa}. The SN ejeta-\otherstar{} interaction is expected to happen naturally in the SD scenario. The thermonuclear explosion ejects the ashes at speeds of order 10$^4\,\mathrm{km\,s^{-1}}$, impacting the \otherstar{} star and forcefully stripping  material away from its surface, changing its structure and composition  \citep{Wheeler1975}. The existence of the \otherstar{} star leads to asymmetric ejecta with a cone-shaped cavity, and the ejecta-\otherstar{} interaction could produce early UV flash \citep{Kasen2010, Liu2015c,Liu2016}. Additionally, the \otherstar{} survives the explosion and should be identifiable by its unusual peculiar velocity. However, searches for such surviving \otherstar{} stars have not yet borne fruit from the standard SD SN Ia channel \citep{Ruiz-Lapuente2004, Ruiz-Lapuente2019, Schaefer2012, Kerzendorf2009, Kerzendorf2014}. 

Many numerical simulations have been constructed to investigate the interaction between SN~Ia ejecta and a stellar \otherstar{} in the SD scenario  \citep[e.g.][]{Marietta2000,Pakmor2008,Pan2010,Pan2012,Liu2012,Liu2013a,Liu2013b,Liu2013c, Boehner2017, Bauer2019,Zeng2020}, using both grid-based and smoothed particle hydrodynamics (SPH) methods. It was found that $\geq0.1\,M_{\sun}$ of hydrogen-rich material is unbound from an H-rich \otherstar{} star in the SD scenario. As a result, the H lines caused by the stripped \otherstar{}-star material are expected to be seen in late-time spectra of SNe~Ia \citep{Mattila2005, Botyanszki2018}. However, no H/He lines have been detected in late-time spectra of normal SNe~Ia \citep[e.g.,][]{Leonard2007, Lundqvist2015, Maguire2016, Tucker2019} and only in two rapidly-declining, subluminous events \citep{Kollmeier2019}. This poses a serious challenge to the SD scenario for SNe~Ia. However, most of previous impact simulations adopt either a simplified structure and composition of the SN ejecta or use the W7~model \citep{Nomoto1984} to represent a SN~Ia explosion. This leads to uncertainty not only in the stripped-mass calculation but also in the predictions of the H-line strengths in late-time spectra of SNe~Ia because the exact explosion mechanism of SNe~Ia is still unknown.

A number of explosion models have been proposed to cover various progenitor scenarios of SNe~Ia \citep{Hillebrandt2013}, including near Chandrasekhar-mass deflagration to detonation transitions (DDT) \citep{Gamezo2005, Roepke2007a, Khokhlov1991, Seitenzahl2013},  Chandrasekhar-mass deflagrations \citep{Nomoto1984, Jordan2012a, Kromer2013, Fink2014}, gravitationally-confined detonations \citep{Jordan2008, Seitenzahl2016},  sub-Chandrasekhar-mass double detonations \citep{Livne1990a, Woosley1994, Fink2007, Kromer2010,Gronow2020}, violent mergers \citep{Pakmor2010, Pakmor2011}, and the collisions of two WDs \citep{Raskin2009,Rosswog2009}. Near Chandrasekhar-mass explosion models in the SD scenario have long been proposed as favourites for SNe~Ia because they well reproduce some observational features such as their light curves and spectra \citep[e.g.,][]{Hoeflich1995, Hoeflich1996, Kasen2009, Blondin2011, Sim2013}. The carbon deflagration model proposed by \citet{Nomoto1984}, known as the W7 model, is commonly used. In this model, the ignition of carbon in the core propagates outwards as a subsonic burning front.  Rayleigh-Taylor instabilities at the contact between hot ashes and cold fuel increase the surface area of the front allowing it to engulf the entire star. More recently, \citet{Seitenzahl2013} proposed a suite of three-dimensional delayed-detonation models of near Chandrasekhar-mass WDs, most notably the N100 model in which 100 ignition sparks begin the explosion deep inside the white dwarf \citep{Roepke2012, Seitenzahl2013}. Alternatively, in the delayed-detonation model, the propagating carbon-ignition flame accelerates to a supersonic velocity, resulting in a pressure wave which causes the surrounding fuel to auto-ignite.

To use observations of type~Ia supernovae to determine their origin, we must compare observations to detailed models including the ejecta-\otherstar{} interaction. Modelling the explosion mechanism correctly is vital to this comparison. Therefore, in this work, we perform two-dimensional hydrodynamic simulations using two widely-studied near Chandrasekhar-mass explosion models within the SD progenitor scenario, i.e., the W7 and N100~models \citep{Nomoto1984,Seitenzahl2013}. We calculate how different explosion mechanisms alter the ejecta-\otherstar{} interaction and expected observational properties, as required to place constraints on the progenitor systems and explosion mechanism of type~Ia supernovae.

\section{Numerical Methods and models}

To investigate type Ia supernova ejecta-\otherstar{} interaction in detail, we use the publicly available hydrodynamics code \textsc{FLASH} version~4.6\footnote{\url{https://flash.rochester.edu/} to solve the Eulerian hydrodynamics equations in 2D cylindrical coordinates}. \textsc{FLASH} is a high-performance, modular, grid-based hydrodynamics framework that could be used in a wide range of physical applications \citep{Fryxell2000}. In our setup, we adopt Helmholtz equation of state \citep{2000ApJS..126..501T} which interpolates thermodynamic solutions from a table of the Helmholtz free energy including contributions from degenerate electrons and positrons, radiation sources and ionised particles. We use adaptive-mesh refinement (AMR) to ensure that small-scale turbulence and instabilities are resolved in locations of interest while computational time is reduced by derefining areas far from the \otherstar{} star. Our grid is a 2D axisymmetric and cylindrical, centered on the \otherstar{}  star, with axes $(r,z)$. 

\subsection{A main-sequence \otherstar{}  model}
\label{sec:companion}

\begin{figure*}
    \centering
    \includegraphics[width=\columnwidth]{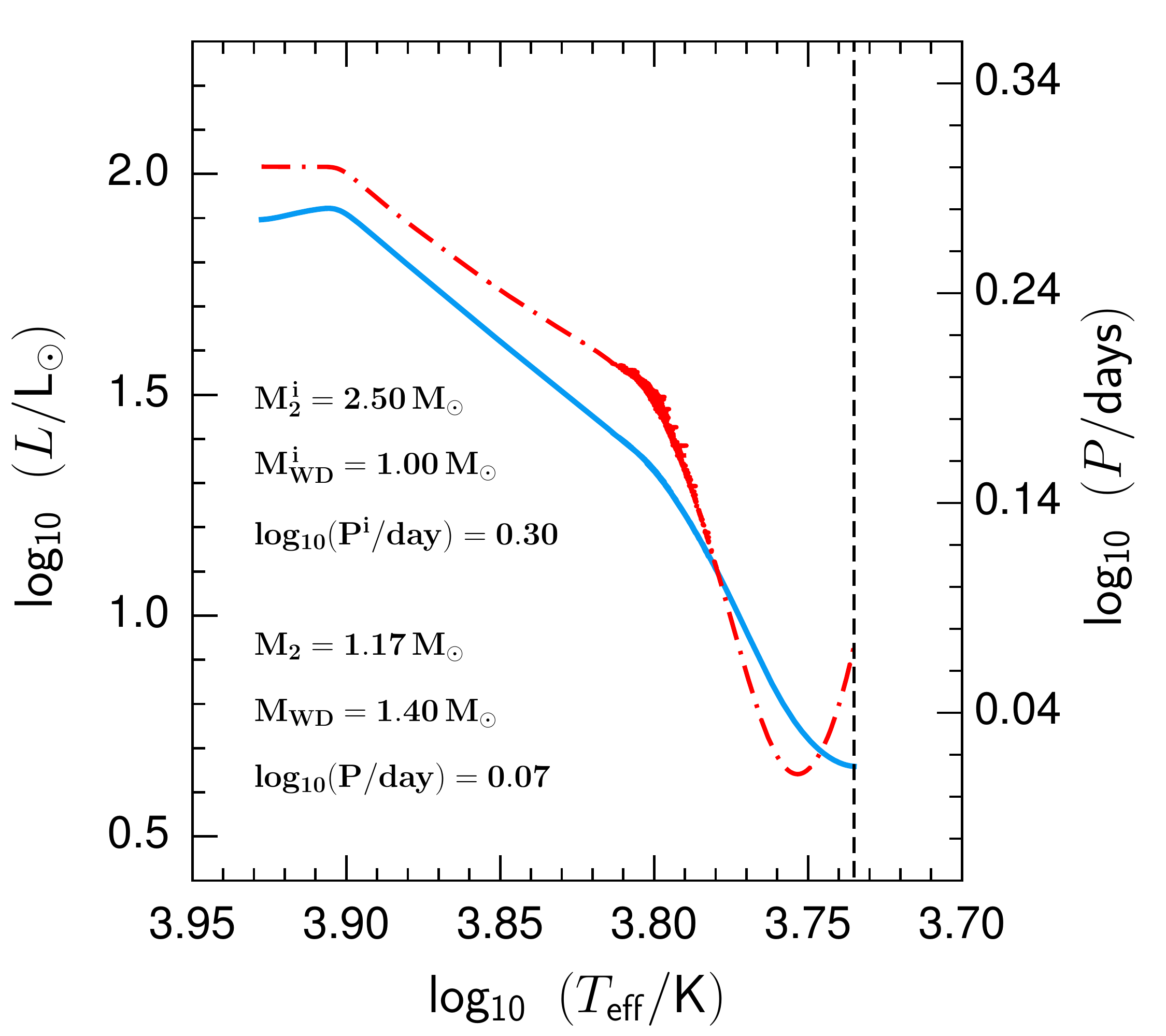}
    \hfill
    \includegraphics[width=\columnwidth]{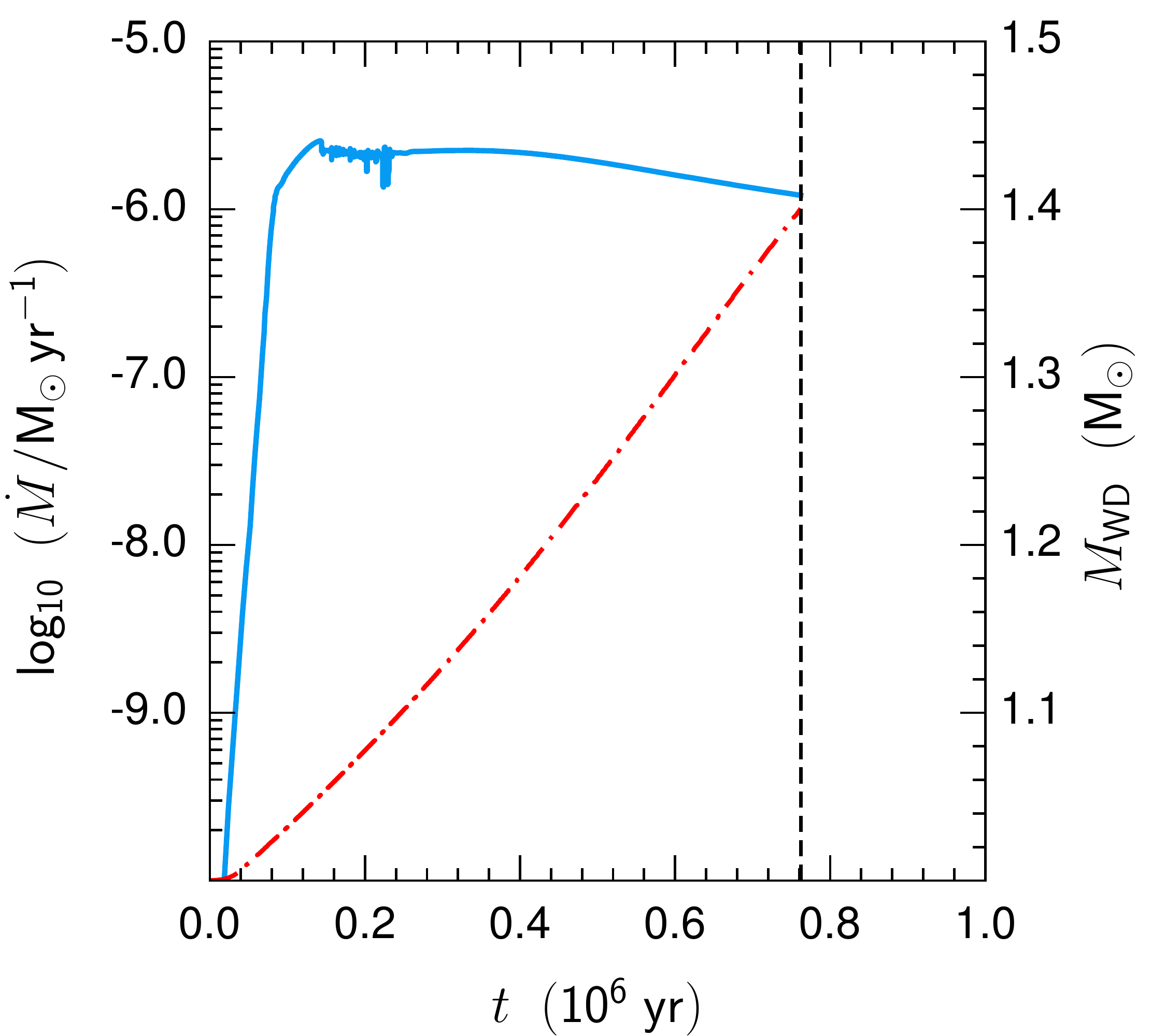}
    \caption{\textit{Left}: Hertzsprung-Russell diagram showing change in luminosity ($L$, solid curve, left axis) and orbital period ($P$, dash-dot curve, right axis) with effective temperature. \textit{Right}: Mass transfer rate ($\dot{M}$, solid curve) and white dwarf mass ($M_{\mathrm{WD}}$, dash-dot curve) vs time since the start of mass transfer. In both figures the dashed vertical line represents the time of SN Ia  explosion.  }
    \label{fig:companion}
\end{figure*} 

\begin{figure}
    \includegraphics[width=\columnwidth]{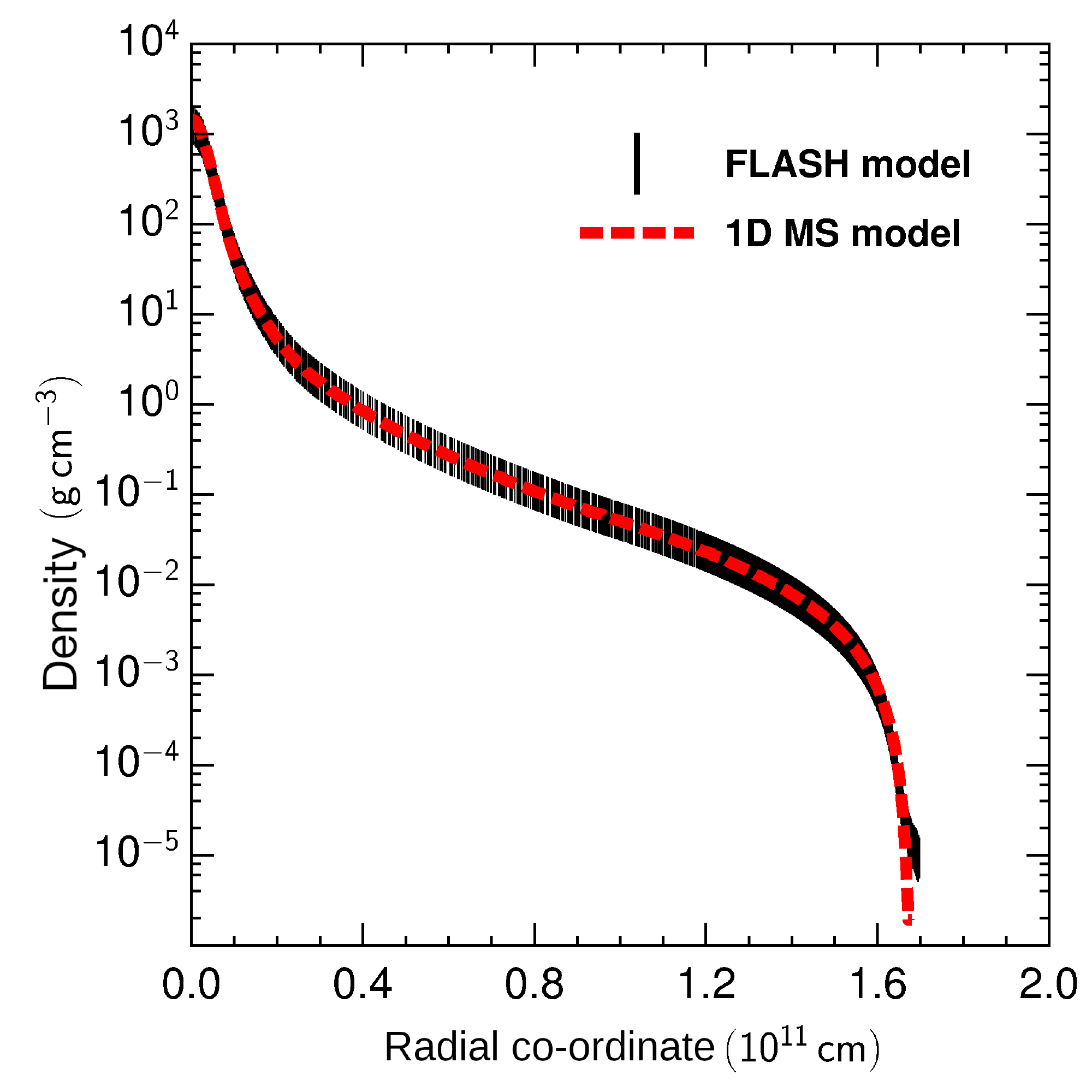}
    \caption{The radial density profile of our 1D \otherstar{} model constructed with \textsc{MESA} (red dashed line) and its corresponding 2D \textsc{FLASH} model (black line) after  relaxation. } 
    \label{fig:ms-dens}
\end{figure}

 The close-binary system initially consists of a MS~\otherstar{} star ($M_\mathrm{2}^\mathrm{i}=2.5\,M_{\sun}$) and a WD~accretor ($M_\mathrm{WD}^\mathrm{i}=1\,M_{\sun}$), with an orbital period  $P_\mathrm{orb}^\mathrm{i}=2\,\mathrm{days}$. This is a typical SN~Ia progenitor system according to the population synthesis studies of \citet{ChenHL2014} and \citet{Claeys2014}.
 To create an initial \otherstar{} star model we use an one-dimensional (1D) stellar evolution code, \textsc{MESA} \citep[version~4906;][]{Paxton2011,Paxton2013,Paxton2015,Paxton2018}. The WD is modelled as a point mass with retention efficiencies of H and He burning from \citet{Yaron2005} and \citet{Kato2004}, respectively. Orbital evolution, including gravitational-wave radiation and associated angular-momentum loss, is modelled as in \citet{ChenHL2014}.

 The binary system is evolved until the WD accretes enough material to reach the Chandrasehkar-mass limit ($\sim 1.4\,M_{\sun}$). We show in Fig.~\ref{fig:companion} the evolution of mass transfer rate and the HR diagram of the \otherstar{} star. At the time of explosion, the binary separation is $A=6.44\,R_{\sun}$, while the \otherstar{} radius and mass are $R_{2}=1.68\times10^{11}\,\mathrm{cm}$ and $M_\mathrm{2}=1.17\,M_{\sun}$, respectively. The density structure of the \otherstar{} star at this moment is shown in Fig.~\ref{fig:ms-dens}.

\subsection{Explosion models}

\begin{figure}
    \includegraphics[width=\columnwidth]{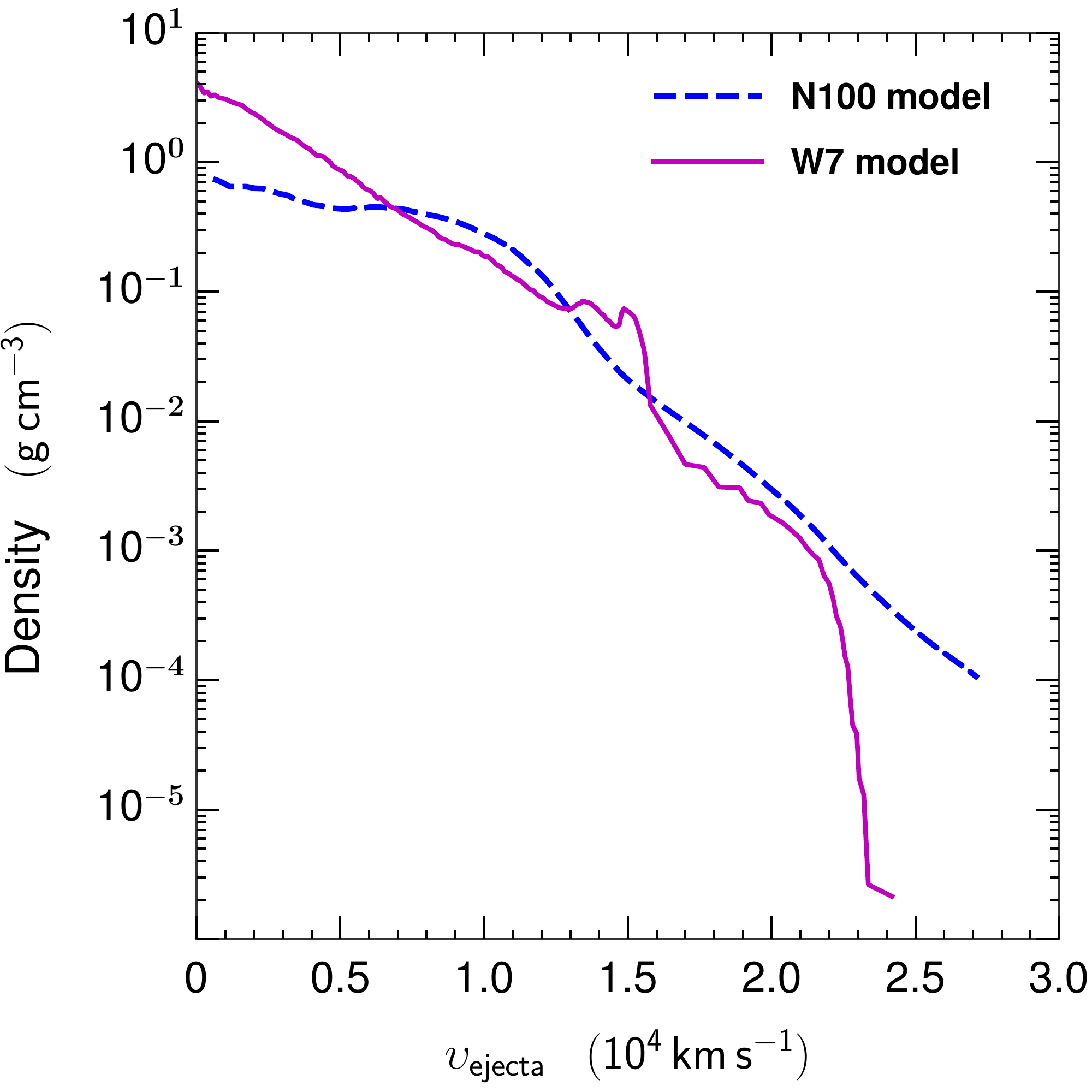}
    \caption{Density as a function of ejecta velocity in the W7 (solid line) and the N100 models (dashed line) about $100\,s$ after the start of the explosion.} 
    \label{fig:sn-dens}
\end{figure}

The W7 carbon deflagration model \citep[][]{Nomoto1984,Maeda2010} \footnote{In particular, the W7 model of \citet{Maeda2010} is used in this work.} and the N100 delayed-detonation model \citep{Seitenzahl2013} are employed in this work. With both, a one-dimensional, angle-averaged shell model of the explosion $20\,s$ after ignition is interpolated onto a two-dimensional grid at the maximum refinement level of our hydrodynamical model to reduce the severity of numerical grid artifacts in the expanding gas front. Both explosions eject $1.4\,M_{\sun}$ but the N100 model has a higher kinetic energy in its ejecta  ($1.40\times10^{51}\,\mathrm{erg}$ compared to $1.23\times10^{51}\,\mathrm{erg}$ in W7). Fig.~\ref{fig:sn-dens} compares the density as a function of expansion velocity in N100 and W7. The N100 model also has a higher-velocity tail in the outer region of ejecta compared to W7 (see Fig.~\ref{fig:sn-dens}).

\begin{figure*}
    \centering
    \includegraphics[width=0.8\textwidth]{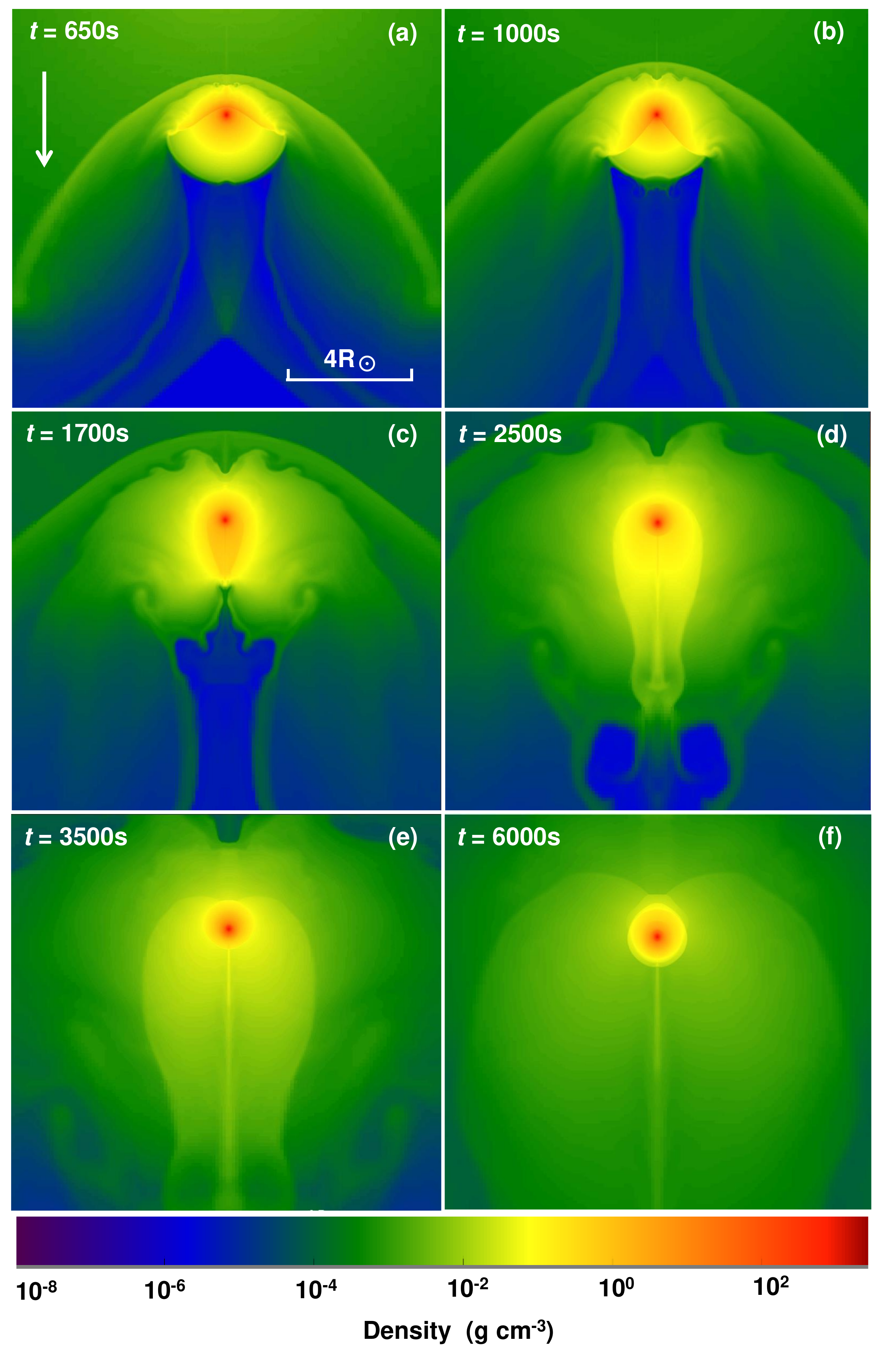}
    \caption{Density slices at times $t\approx650\,\mathrm{s}$, $1000\,\mathrm{s}$, $1700\,\mathrm{s}$, $2500\,\mathrm{s}$, $3500\,\mathrm{s}$ and $6000\,\mathrm{s}$ after explosion in our 2D hydrodynamical simulation of the N100 model \citep{Seitenzahl2013}. The colour scale shows the logarithm of the mass density. The direction of motion of the incoming SN ejecta is from top to bottom (see arrow symbol). The plots are made  using \texttt{yt} \citep[][]{Turk2011}.}
    \label{fig:n100-all}
\end{figure*}

\begin{figure*}
    \centering
    \includegraphics[width=0.86\textwidth]{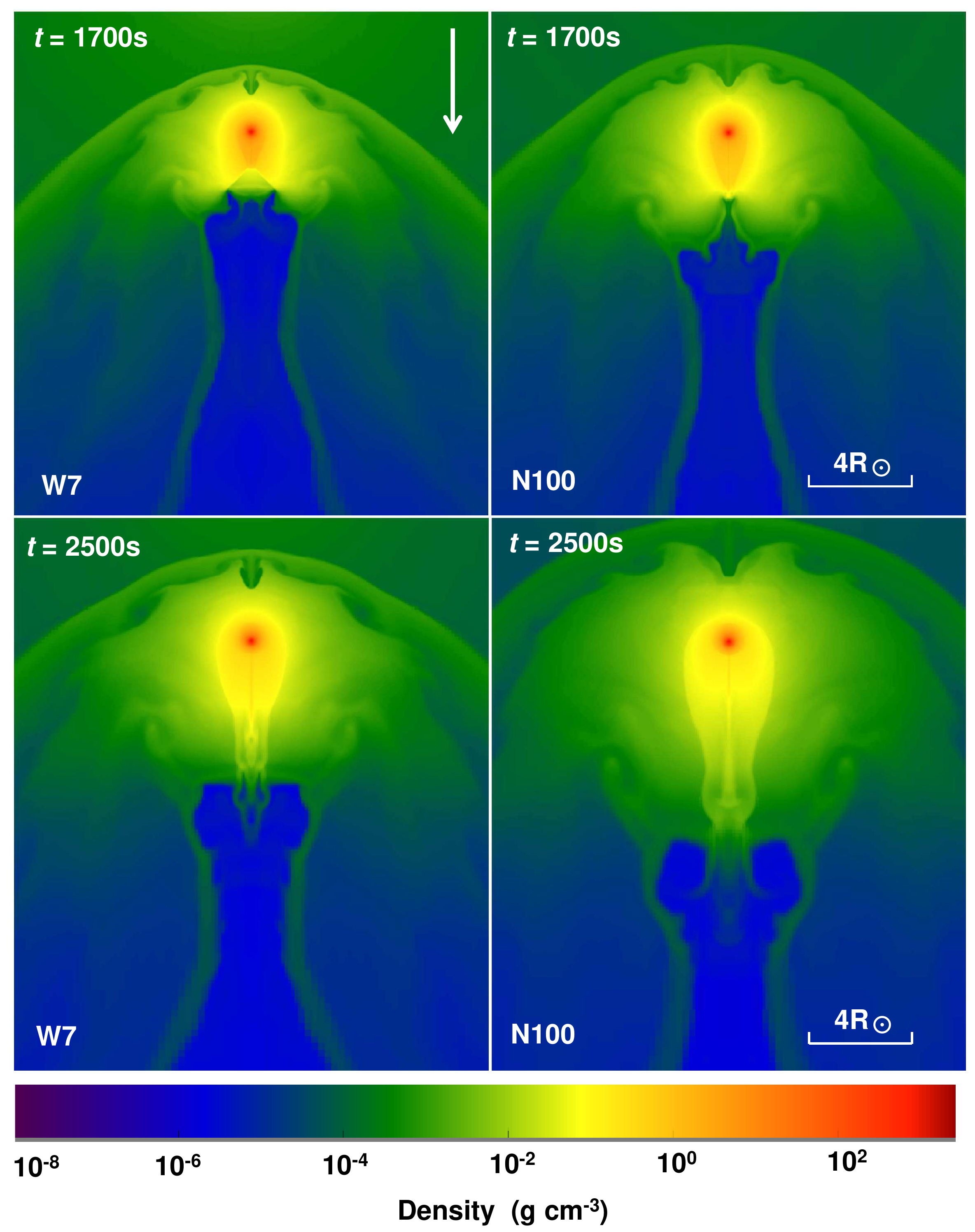}
    \caption{A comparison of density slices at $t\approx1700\,\mathrm{s}$, $2500\,\mathrm{s}$, between the W7 model (left column) and the N100 model (right column) from our 2D impact simulations. The color scale shows the logarithm of the mass density. The direction of motion of the incoming SN ejecta is from top to bottom (see arrow symbol). The plots are made by using \texttt{yt} \citep[][]{Turk2011}. }
    \label{fig:dens500}
\end{figure*}

\subsection{Setup and key assumptions}

The simulation setup is similar to what has been described in \cite{Pan2010, Pan2012} but upgraded to FLASH~4 version. To initialise the hydrodynamic simulation, the MS \otherstar{}-star model is interpolated onto a two-dimensional grid from the one-dimensional shell model provided by MESA. The system is then relaxed for 10$^{4}\,\mathrm{s}$ which artificially damps the momentum of the star so it settles into hydrostatic equilibrium. Next, the W7 or N100 SN~Ia explosion is placed at a distance $A$ from the centre of the \otherstar{} star on the $z$-axis. The simulation then runs for a further 10$^{4}\,\mathrm{s}$ until the ejecta-\otherstar{} interaction finishes and the mass of the \otherstar{} star stabilises. Because the \otherstar{} fills its Roche lobe at the moment of supernova, the binary separation ($A$) and the \otherstar{} radius ($R_{2}$) follow Equation~\ref{eq:eggletonsep} \citep{Eggleton1983},
\begin{equation}
    \frac{R_{2}}{A} = \frac{0.49q^{2/3}}{0.6q^{2/3}+\ln{\left(1+q^{1/3}\right)}}\,, 
    \label{eq:eggletonsep}
\end{equation}
where $q$ is the mass ratio of binary system at the moment of explosion.

In our simulations we employ 12~levels of adaptive-mesh refinement (AMR) in the SN and \otherstar{}-star regions, and a maximum of 10~levels of refinement outside these regions. We refer to this setup as the 10/12~level. This is equivalent to a uniform grid spacing of $6.1\times10^{7}\,\mathrm{cm}$ ($\sim3.6\times10^{-4}\,R_{2}$) inside the star and supernova regions, and a maximum of $9.8\times10^{6}\,\mathrm{cm}$ ($\sim5.8\times10^{-3}\,R_{2}$) outside of these regions.  The cylindrical coordinates $r,z$ lie in the ranges  $r=(0, 2\times10^{12})$ and $z=(-2\times10^{12}, 2\times10^{12})$, which are about 12 times larger than the \otherstar{}-star radius in each direction.

To determine the robustness of numerical results, \citet{Pan2010} performed a convergence test using different maximum AMR levels from 6/8 to 10/12 in their 2D impact simulations with \textsc{FLASH}. They found that the differences in the results, e.g.~to the unbound mass and kick velocity, given by these different resolution levels are not significant, and that level 10/12 is sufficient to study ejecta-\otherstar{}-star interaction. In addition, the convergence tests of \citet{Pan2010} were performed with a helium-star companion which is more compact than our main-sequence donor. The turbulence generated from ejecta-companion impact is weaker in the main-sequence channel and we expect better convergence with main-sequence companions. Because the \textsc{FLASH} code is used in this work, albeit a newer version, and our initial setup and assumptions are similar to those of \citet{Pan2010}, we use the recommended 10/12 level in all our simulations.

We neglect orbital and rotational motion in our simulations because the SN~Ia ejecta velocity ($\sim10^{4}\,\mathrm{km\,s^{-1}}$) far exceeds the orbital velocity (a few $\times 10^{2}\,\mathrm{km\,s^{-1}}$), and the two stars are tidally locked at the time of explosion so the \otherstar{} surface also rotates at the orbital velocity.
 
\begin{figure}
   \includegraphics[width=\columnwidth]{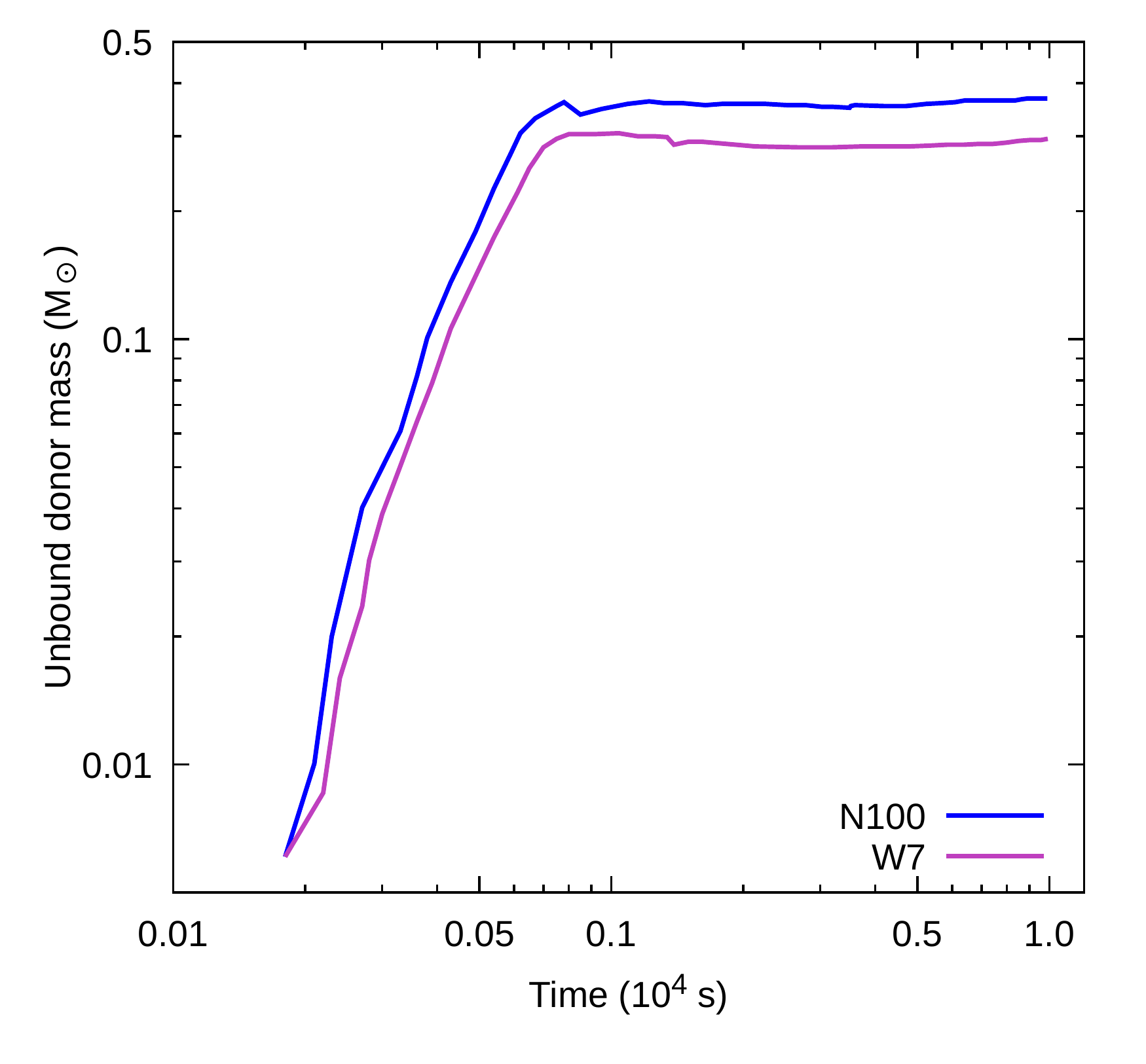}
    \caption{Unbound \otherstar{}-star mass vs time since the supernova in our simulations with the W7 (dash-dotted line) and N100 (solid line) explosion models. Most mass is lost from the \otherstar{} just after the blast wave shocks its surface, before about $100\,\mathrm{s}$, with later thermal relaxation causing a small amount of mass to re-bind by the end of the simulation.}
    \label{fig:ubmass}
\end{figure}

\begin{figure}
   \centering
    \includegraphics[width=\columnwidth]{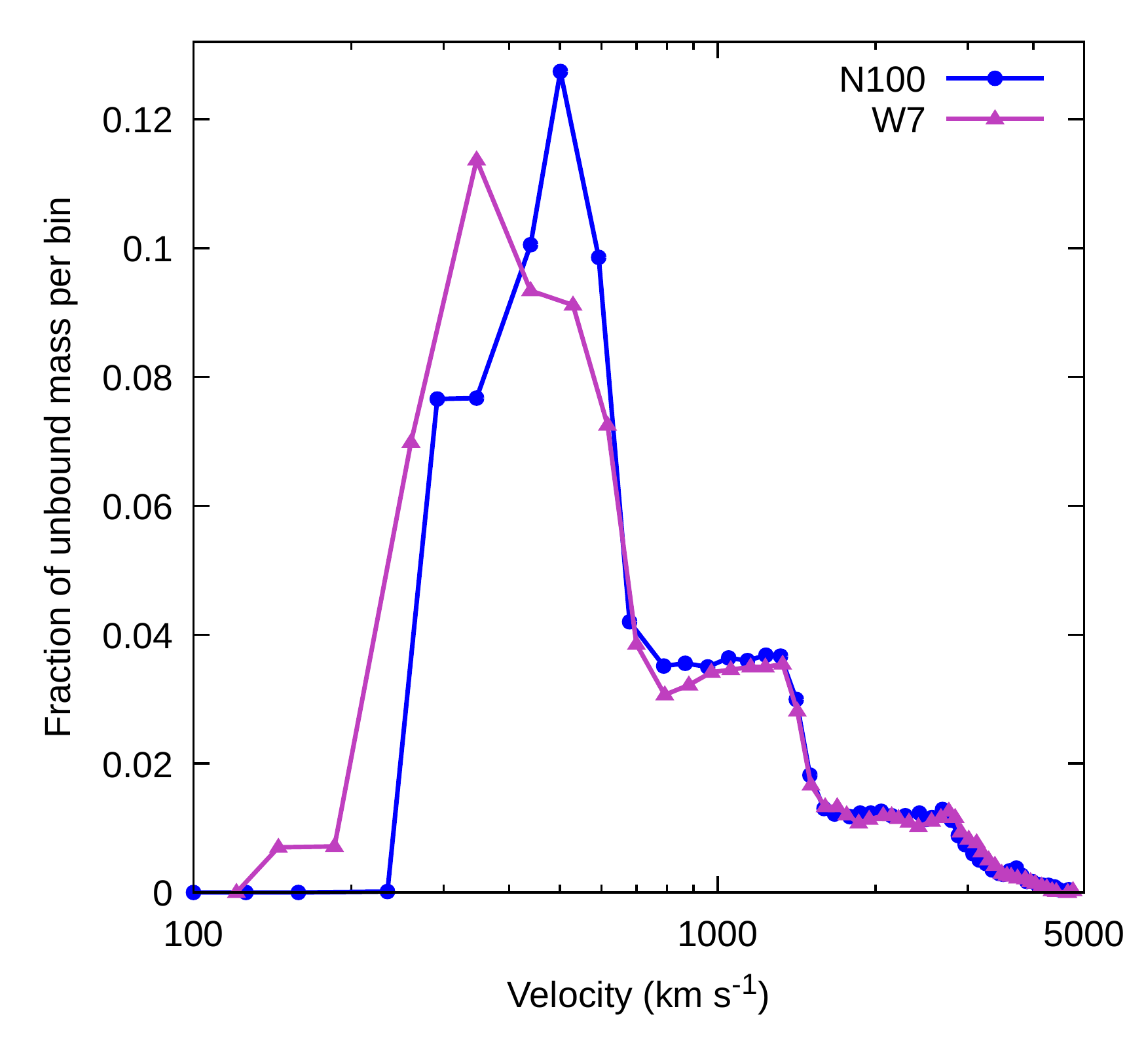}
    \caption{Velocity distributions of unbound \otherstar{}-star material at end of our simulations using the two different explosion models W7 (blue dashed top-left to bottom-right) and N100 (magenta dashed bottom-left to top-right). }
    \label{fig:velhist}
\end{figure}

\section{Results}

In this section, we present the results of our 2D hydrodynmical simulations of ejecta-companion interaction, including the stripped mass and kick velocity. A comparison of the results between W7 model and N100 model is given to discuss the effect of different explosion models on the details of ejecta-companion interaction.

\subsection{Ejecta-\otherstar{} interaction}

Fig.~\ref{fig:n100-all} shows the evolution of the ejecta-\otherstar{} interaction in our N100 model. The exploding star is located vertically above the figure, out of shot, and the blast wave moves downwards towards the \otherstar{}.
The SN~ejecta expands freely for a few minutes then hits the surface of the \otherstar{} star. The impact strips some H-rich material from and increases the pressure on the surface of the \otherstar{}. A high-pressure shock is driven into the stellar interior and a bow shock develops in the outer regions which diverts SN~ejecta material around the star. As the diverted material further expands around the \otherstar{} star, the interstellar medium (ISM) is compressed in the downstream region (below the star in the figure). The two opposing fronts collide behind the star at a few hundred seconds (Fig.~\ref{fig:n100-all}a), forcing ISM material back towards the star, whose impact can be observed as a small notch at its rear. The shock continues to pass through the \otherstar{} star, slowing as it moves through the higher density core (Fig.~\ref{fig:n100-all}b). The opening hole is created behind the \otherstar{} star.

Instabilities develop on the \otherstar{} surface closest to the SN, disrupting its initially-spherical shape. At about $t=1700\,\mathrm{s}$ the shock passes through the entire star and pushes stellar material out of its rear as it exits (Fig.~\ref{fig:n100-all}c). The symmetries of the \otherstar{} star are completely destroyed as the shock passes through it. The stripped \otherstar{} material  mixes with the SN ejecta material in the opening hole behind the star (Fig.~\ref{fig:n100-all}d). The outer layers of the star expand in the shape of an inverted teardrop, surrounded by  Rayleigh-Taylor instabilities. Instabilities are also  observed at the edge closest to the SN, travelling outwards from the centre of the \otherstar{} but shielded from incoming material by a bow shock. The \otherstar{} star expands because of the shock heating, and is totally out of hydrostatic and thermal equilibrium. By $t=6000\,\mathrm{s}$, the \otherstar{}  starts returns to hydrostatic equilibrium as the interaction finishes and a relatively spherical remnant remains (Fig.~\ref{fig:n100-all}f). At the same time, an expanded low-density envelope has forms in the outer region of the star. In addition, the centre of the \otherstar{} star moves away from its original position because it receives a velocity kick because of momentum imparted by the SN ejecta during the interaction.

\begin{table*}\renewcommand{\arraystretch}{1.4}
\fontsize{9}{11}\selectfont
\caption{Parameters summarising our initial models and results. \label{table:1}}
\centering
\begin{tabular}{ccccccccccc}    
\hline\hline
Simulation & SN model & $E_\mathrm{kin,SN}$ & $M_{\rm{WD}}^\mathrm{i}$ & $M_{\rm{2}}^\mathrm{i}$ & $M_{\rm{2}}$ & $R_{\rm{2}}$ & $A$  & $\Delta M_{\mathrm{stripped}}$ & $V_{\rm{kick}}$ & $V_{\mathrm{modal}}$\\

& & $(\mathrm{10^{51}\,erg})$& $(\mathrm{M}_{\odot})$ & $(\mathrm{M}_{\odot})$ & $(\mathrm{M}_{\odot})$ &$(\mathrm{10^{11}\,cm})$& $(\mathrm{10^{11}\,cm})$ & $(\mathrm{M}_{\odot})$ & $(\mathrm{km\,s^{-1}})$ & $(\mathrm{km\,s^{-1}})$ \\ 
\hline
A & W7   & 1.23 & 1.000 & 2.500 & 1.170  & 1.68 & 4.48 &  0.293 & 73.8 & 350\\
B & N100 & 1.40 & 1.000 & 2.500 & 1.170  & 1.68 & 4.48 &  0.368 & 86.0 & 500\\
\hline
\end{tabular}
\flushleft
$E_\mathrm{kin,SN}$ is the kinetic energy of SN ejecta of the explosion model. $M_{\rm{WD}}^\mathrm{i}$ and $M_{\rm{2}}^\mathrm{i}$ are the WD and \otherstar{} masses at the beginning of mass transfer. $M_{\mathrm{2}}$, $R_{\rm{2}}$ and $A$ are the \otherstar{} mass, \otherstar{} radius and binary separation at the time of  explosion. $\Delta M_{\mathrm{stripped}}$ and $V_{\rm{kick}}$ are the mass stripped from the \otherstar{} and the kick velocity received by the \otherstar{} star at the end of our 2D impact simulations. $V_{\rm{modal}}$ is the typical velocity of stripped \otherstar{} material (Fig.~\ref{fig:velhist}).
\end{table*}

\subsection{Stripped mass and kick velocity}

A comparison of the interaction between the W7 and N100 explosion models at equal times after the supernova explosion is shown in Fig.~\ref{fig:dens500}. Both explosion models lead to similar interaction characteristics: a turbulent front behind the frontal bow shock, eddies to the sides of the rear of the star, and the teardrop-shaped plume ejected after the shock passes through the star in its entirety. 
The clearest difference is that the N100 explosion has faster ejecta, hence the interaction evolves more quickly than with  W7 and strips more material from the \otherstar{} star.

Fig.~\ref{fig:ubmass} shows the unbound \otherstar{} mass as a function of time using the two explosion models in our simulations. To determine whether the \otherstar{} material is unbound, we calculate its total energy by summing the kinetic, potential and internal energies, $E_\mathrm{tot}=E_\mathrm{kin}+E_\mathrm{pot}+E_\mathrm{int}$, and if $E_\mathrm{tot}>0$ the material is unbound. As shown in Fig.~\ref{fig:ubmass}, the unbound \otherstar{} mass stabilises about $5000\,\mathrm{s}$ after the explosion. The time-evolution of unbound mass is similar in both models, beginning with a steep rise as the shock front transfers momentum and thermal energy to the \otherstar{} star, followed by a slight decrease during  thermal relaxation which returns some material to a bound state. At the end of our simulations, we find that $0.293\,M_{\sun}$ and $0.368\,M_{\sun}$ of H-rich material, representing $25$ and $31\,\mathrm{per~cent}$ of the \otherstar{} mass at the time of explosion, is stripped from the MS \otherstar{} surface by the SN explosion using the W7 and N100 models respectively.

Extra mass stripping in our N100 model is caused by the larger kinetic energy of the N100 explosion, $1.40\times10^{51}\,\mathrm{erg}$ \citep{Seitenzahl2013}, compared with W7 which has $1.23\times10^{51}\,\mathrm{erg}$ \citep{Nomoto1984}. Previous studies have found that more powerful ejecta generally strip more \otherstar{} mass for a given \otherstar{} model and binary separation, and the total stripped mass is linearly proportional to the SN kinetic energy \citep{Pakmor2008, Pan2012,Liu2013a}. The stripped mass in our W7 model increases to $0.334\,M_{\sun}$ if we scale it with a higher kinetic energy of $1.40\times10^{51}\,\mathrm{erg}$. On the other hand, as shown in Fig.~\ref{fig:sn-dens}, the outer SN ejecta of the N100 model is more powerful than that of W7, hence N100 is also expected to strip more.

As described in Section~\ref{sec:introduction}, the interaction between SN~Ia ejecta and a \otherstar{} star has been investigated by  previous studies with both analytical methods, and two- and three-dimensional hydrodynamical simulations \citep[e.g.,][]{Wheeler1975, Marietta2000,Pakmor2008,Pan2012,Liu2012,Boehner2017}. Most studies found that $\gtrsim 0.1\,M_{\sun}$ can be stripped from a main-sequence \otherstar{} star, from a red giant the whole envelope is stripped. In particular, \citet{Marietta2000} found that $15\,\mathrm{per~cent}$ of the \otherstar{} star is stripped in their HCV model. The HCV model has a mass $M_{2}=1.017\,M_{\sun}$, radius $R_{2}=6.8\times10^{10}\,\mathrm{cm}$ and ratio of binary separation to radius $A/R_{2}=3.0$ \citep{Marietta2000}. Compared to our main-sequence donor model, the HCV model is more compact and has a larger $A/R_{2}$ (Table~\ref{table:1}). The stripped mass in HCV increases to $\sim0.2\,M_{\sun}$ \citep[][see their Fig.~12]{Pan2012} if we reduce its $A/R_{2}$ to 2.67 similarly to our model. \citet{Liu2012} found between 6 and 24 per cent depending on the MS \otherstar{} model \citep[see also][]{Pan2012}. More recently, \citet{Boehner2017} predicted $21-26\,\mathrm{per~cent}$ of a main-sequence \otherstar{} is stripped in their 2D simulations using \textsc{FLASH}. Our results are comparable with these previous studies. As shown in Figs.~\ref{fig:ms-dens} and~\ref{fig:companion}, our \otherstar star is near the end of its main sequence at the time of SN explosion, hence it has a dense core with an extended envelope. This explains why our stripped mass is a greater than in \citet{Boehner2017}, in which the donor stars are normal MS stars and they have a more compact envelope than our donor model. However, our stripped mass greatly exceeds the $\sim1$--$5\,\mathrm{per~cent}$ predictions of \citet[][their Table~2]{Pakmor2008}. The differences are mainly caused by the detailed structure of the \otherstar{} star models and binary separations at the moment of SN Ia explosion \citep[][]{Liu2012,Pan2012,Boehner2017}.

Compared to our model, the main-sequence \otherstar{} models of \citet{Pakmor2008} are more compact and have longer binary separations (i.e., they have a larger $A/R_{2}$) so in their simulations less mass is stripped. For instance, model rp3\_24a in \citet{Pakmor2008} has a binary separation of $4.39\times10^{11}\,\mathrm{cm}$ which is similar to our $4.48\times10^{11}\,\mathrm{cm}$. However, our companion radius exceeds theirs ($7.21\times10^{10}\,\mathrm{cm}$) by a factor of $2.3$. Additionally, model rp3\_20a in \citet{Pakmor2008} has the same mass as ours (i.e., $M_{2}$=$1.17\,M_{\odot}$), but their $A/R_{2}=4.5$ exceeds ours ($A/R_{2}=2.67$) by a factor of 1.7. If we assume that their rp3\_20a model also has $A/R_{2}=2.67$, their stripped mass increases to $0.19\,M_{\odot}$ based on their Equation~4. The main differences between the main-sequence \otherstar{} models of \citet{Pakmor2008} and ours are the different treatments of pre-SN mass-loss of the MS stars and the binary separations adopted at the moment of SN explosion, which have been discussed in detail in our previous work  \citep[][their Section~4.1]{Liu2012}. \citet{Pakmor2008} start the mass accretion phase with less-evolved companion stars and they constructed their companion star models by rapidly removing mass while evolving a single main-sequence star to mimic pre-SN mass-transfer in a binary system. As a consequence, their models have a very large $A/R_{2}$, and they are too compact to fill their Roche-lobe at the moment of SN explosion. In contrast, our main-sequence donor model is constructed by consistently following mass-transfer in a binary system. Our main-sequence star has a more extended structure and fills its Roche-lobe when the SN explodes.

The \otherstar{} star receives a kick as a result of momentum transfer from the ejecta during the interaction. The time evolution of kick velocity behaves similarly to the stripped mass (Fig.~\ref{fig:ubmass}).  At the end of simulations ($t=10^{4}\,\mathrm{s}$), the kick velocities are $73.8\,\mathrm{km\,s^{-1}}$ and $86.0\,\mathrm{km\,s^{-1}}$ in our W7 and N100 models respectively. The faster kick in the N100 model is probably because its ejecta have a higher kinetic energy than in W7. The movement of the \otherstar  away from its original position is seen in Fig.~\ref{fig:n100-all}.
The final spatial velocity of the surviving \otherstar{} star is only slightly altered by the kick, with most of the velocity arising from its pre-explosion orbital motion.

\begin{figure*}
    \includegraphics[width=\columnwidth]{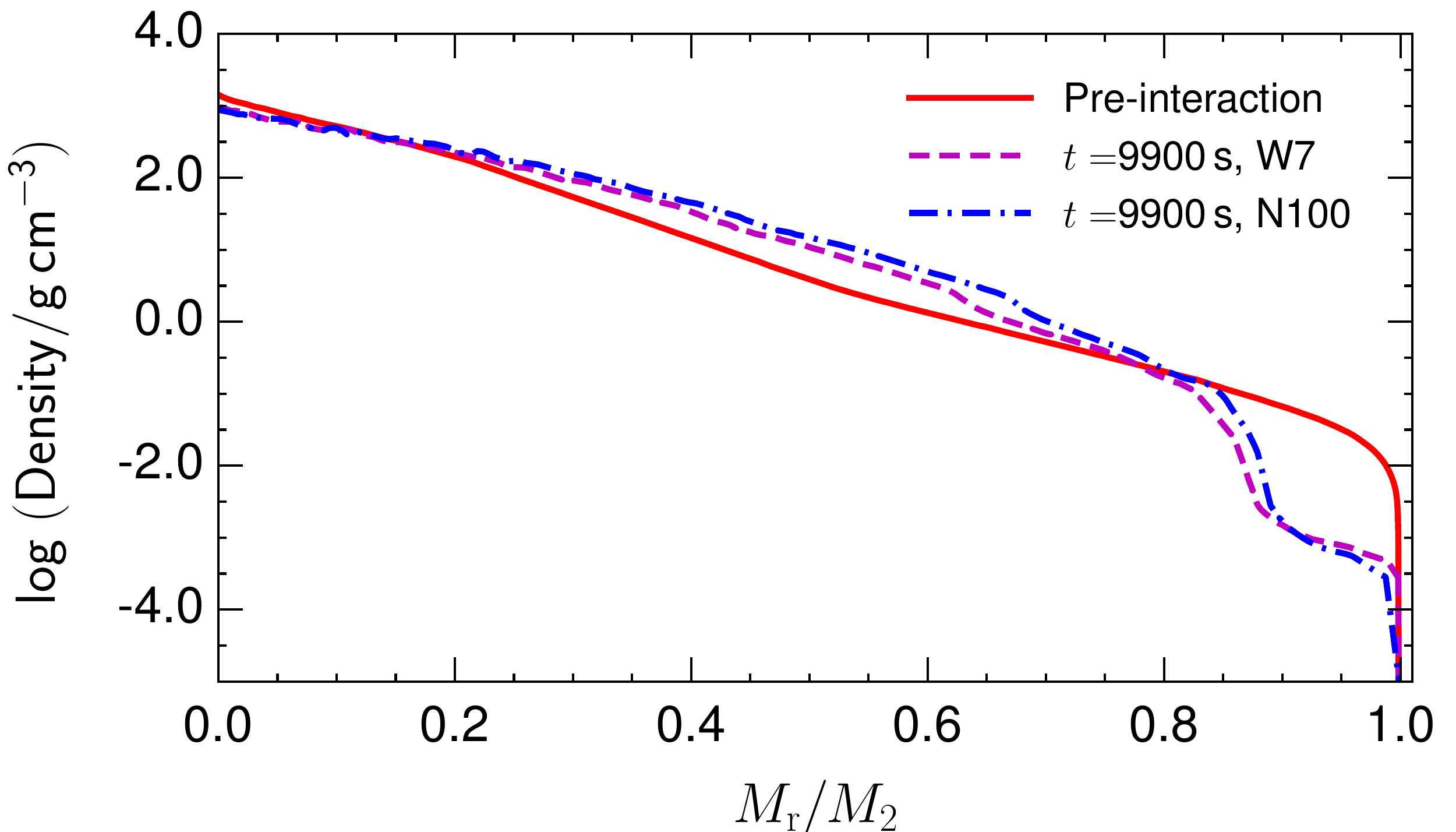}
    \hfill
    \includegraphics[width=\columnwidth]{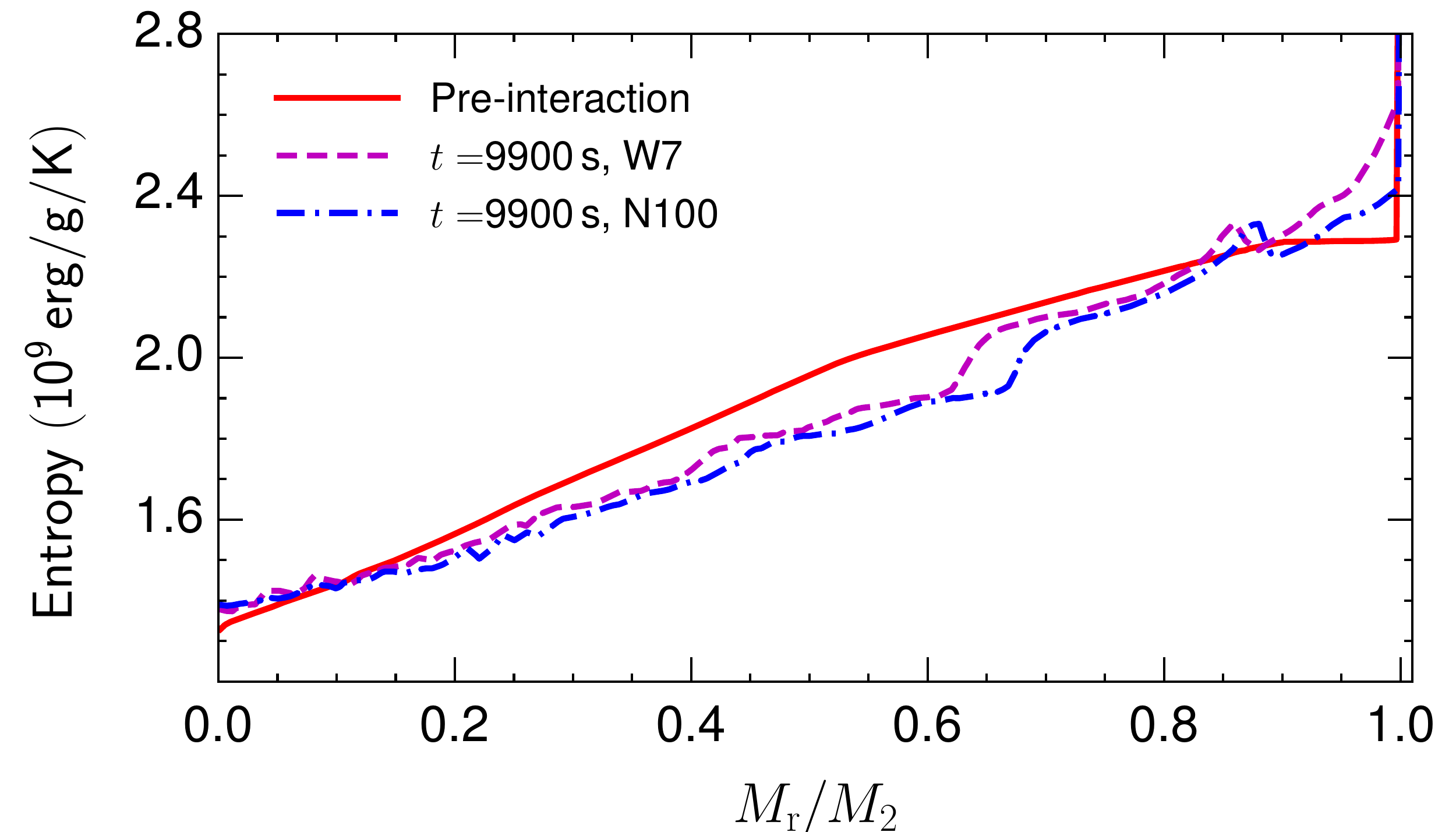}
    \caption{Post-impact 1D averaged density (left panel) and entropy (right panel) of the surviving \otherstar{} star as a function of relative mass, $M_r/M_2$ where $M_r$ is the Lagrangian mass co-ordinate, at the end of our simulations, when $t\approx 10^{4}\,\mathrm{s}$, with the W7 (red dashed line) and N100 (green dash-dotted line) explosion models. For comparison, the corresponding 1D profiles of initial pre-impact \otherstar{} star are also shown as solid lines. }
      \label{fig:survive-profiles}
\end{figure*}

\section{Discussion}

{In this section, we compare our results with the late-time observations of SNe~Ia to examine the validity of the SD progenitor model. In addition, we discuss the distributions of stripped companion material and the properties of surviving companion stars to derive some clues to the identification of SD progenitor model. Uncertainties in our results and future work are also given.}

\subsection{Models compared to observations}

Hydrogen-emission lines will be seen in late-time spectra of SNe~Ia, {about a few hundred days after the explosion},  if sufficient hydrogen-rich material ($>0.001-0.058\,M_{\sun}$, which depends on the distance of the observed SNe Ia) is stripped from the \otherstar{} surface during the interaction of the explosion with the star \citep[][]{Mattila2005,Lundqvist2015,Botyanszki2018,Dessart2020}. 
Therefore, searching for hydrogen-emission lines in late-time spectra of SNe~Ia should provide strong evidence in favour, or otherwise, of a SD progenitor system to observed Ia supernovae. {To date, no hydrogen-emission lines have been detected in late-time spectra of normal SNe~Ia \citep[][]{Leonard2007,Lundqvist2013,Lundqvist2015,Maguire2016,Shappee2016,Tucker2020}. However, hydrogen features have been detected in late-time spectra of two fast-declining, sub-luminous SNe Ia, SN~2018fhw \citep{Kollmeier2019} and SN~2018cqj \citep[][]{Prieto2020}. All this late-time observations limit the hydrogen-rich mass in type-Ia SN progenitor systems to $\leq0.001-0.058\,M_{\sun}$.}

This observational upper limit to the stripped mass is less than our simulations predict, $\geq 0.30\,M_{\sun}$, and our results are similar to previous studies  \citep[][]{Marietta2000,Pakmor2008,Pan2012,Liu2012,Liu2017,Boehner2017}. This suggests that the non-detection of hydrogen features in late-time spectra of SNe~Ia is seriously challenges the SD scenario if we assume that our current hydrodynamical and radiative transfer models of late-time hydrogen emission are accurate.

\subsection{Velocity distributions of stripped material}

The distribution of velocities of  H/He-rich material stripped from the  \otherstar{} star hints at expected  H- and He-emission line features in late-time spectra of SNe Ia  \citep[][]{Mattila2005, Botyanszki2018,Dessart2020}. In Fig.~\ref{fig:velhist}, we show post-impact velocity distributions of stripped material from the surface of our \otherstar{} star in our two explosion models. The typical velocities of stripped material in our W7 and N100 models are $350\,\mathrm{km\,s^{-1}}$ and $500\,\mathrm{km\,s^{-1}}$, respectively, which are consistent with  previous works \citep[][]{Marietta2000,Pan2012,Liu2012}.

Additionally, \citet{Marietta2000} suggests that the high-velocity tail of stripped material is an important diagnostic that allows discrimination between various \otherstar{} models \citep[see also][]{Boehner2017}. Here, we check whether the two explosion models investigated in this work can be discriminated similarly.  Fig.~\ref{fig:velhist} shows that although a high-velocity, $\gtrsim 1000\,\mathrm{km\,s^{-1}}$, tail is seen in both our explosions models, there is little difference between their velocity distributions. This suggests that it may be difficult to discriminate between the W7 and N100 models with the features caused by the high-velocity tail for a given main-sequence \otherstar{} star model.

\subsection{Surviving \otherstar{} stars}

In the SD SNIa scenario, the \otherstar{} star survives the explosion after ejecta-\otherstar{} interaction. However, surviving \otherstar{} stars are generally not expected in the double-degenerate scenario, although \citet{Shen2018} have recently suggested that a WD \otherstar{} may survive in their D6 double-generate model. Therefore, searching for surviving \otherstar{} stars in SN remnants (SNRs) is a promising way to distinguish between the SD and DD  scenarios of SNe Ia \citep[][]{Ruiz-Lapuente2004, Ruiz-Lapuente2019, Kerzendorf2009, Kerzendorf2013,Schaefer2012, Pan2014,Li2019}.

To provide the predicted observational features of surviving \otherstar{} stars for comparison with observations of SNRs, it is important to address long-term post-impact evolution of surviving \otherstar{} stars based on the results of multi-dimensional hydrodynamical impact simulations.  After the explosion, the surviving \otherstar{} star becomes a runaway star. It escapes from the binary system with a spatial velocity caused mostly by its pre-explosion orbital motion. In Figure~\ref{fig:survive-profiles}, we present 1D-averaged density (left panel) and entropy (right panel) profiles of our surviving main-sequence \otherstar{} star at the end of our simulations with the W7 and N100 explosions. For comparison, the initial profiles of the \otherstar{} star at the beginning of the simulations are also shown.  The central density of the \otherstar{} star  decreases by about $50\,\mathrm{per~cent}$ because of the shock that passes through the star. {In addition, the \otherstar{} star is  impacted and heated during the interaction, and expands by a factor of about two by the end of our simulations.}

{While the \otherstar{} subsequently thermally relaxes, it is overluminous compared to its pre-explosion equilibrium state  \citep[][]{Podsiadlowski2003,Pan2012b,Pan2013,Shappee2013,Liu2020,Liu2021,Liu2022ApJ}. For instance, \citet{Pan2012} and \citet{Shappee2013} suggest that the post-impact main-sequence companion stars are significantly more luminous (up to about  $10$--$1000\,L_{\odot}$) within a few thousand years, depending on the main-sequence companion model.} Using the approach of \citet{Liu2020}, the 2D surviving \otherstar{} model computed from our impact simulations in this work can be mapped into a 1D stellar evolution code such as \textsc{MESA} to follow their subsequent post-impact evolution for more than a few hundred years to obtain  expected observational features \citep[][]{Pan2013,Liu2021,Liu2022ApJ}. {This is important to understand how different explosion models affect the observational features of the surviving main-sequence companion stars, and is  useful in the search for such events in SN~Ia remnants.} This investigation will be performed in a forthcoming study.

\subsection{Uncertainties and future work}

Our simulations are performed in 2D without considering the orbital motion and rotation of the binary system. This is expected to only slightly affect our results \citep[][]{Liu2012, Pan2012} because these two velocities are slower than the typical ejecta velocity of $10^{4}\,\mathrm{km\,s^{-1}}$ by more than an order of magnitude. \citet{Pan2012} have performed the ejecta-donor interaction in 2D and 3D for a given main-sequence donor model with the FLASH code. They find that the results of the ejecta-donor interaction are quite consistent in 2D and 3D if the orbital motion and spin of the star are not included. However, the orbital motion and spin can lead to slightly different results in 3D \citep[][their Table~2]{Pan2012}. Future 3D simulations are still needed to make more strict theoretical predictions such as post-SN rotation features of the surviving \otherstar{} star. For instance, exploring the post-impact spin features of the surviving donor star requires the impact simulation to be in 3D.

In addition, the rotation of an accreting WD is not considered when we construct the \otherstar{} model at the moment of SN explosion in our 1D binary evolution calculation. The accretion from the \otherstar{} star spins up the WD, leading to a delay to spin down the WD before it explodes (i.e., the so-called ``spin up/spin down'' model, \citealt{Justham2011, DiStefano2011}). Because the spin-down timescale is quite uncertain, the \otherstar{} star in the MS \otherstar{} scenario could be any type of star from MS (short delay) to WD (long delay) at the time of explosion. In such a case, the amount of stripped \otherstar{} mass and \otherstar{} properties will very likely differ considerably from our results with a MS donor.

\section{Summary and conclusions}

In this work, the effects of different explosions on the results of the ejecta-\otherstar{} interaction in normal SNe~Ia are investigated with 2D hydrodynamical simulations using the \textsc{FLASH} code for a given MS \otherstar{} which is computed by the \textsc{MESA} code. Specifically, two near-Chandrasekhar mass explosion models,  W7 \citep{Nomoto1984} and N100  \citep{Roepke2012, Seitenzahl2013}, are adopted to represent the SN~Ia explosion in our simulations. The main goal of this work is to investigate how different explosion models affect the results of ejecta-\otherstar{} interaction such as the total amount of stripped \otherstar{} material and the kick velocity received by the \otherstar{} star during the interaction. We find that  $0.30\,M_{\sun}$ and $0.37\,M_{\sun}$ are stripped from a typical $2.5\,\mathrm{M}_{\sun}$ main-sequence \otherstar{} surface by the SN~Ia explosion, in simulations with the W7 and N100 explosion models, respectively. Additionally, the \otherstar{} star  receives a kick velocity of about $74\,\mathrm{km\,s^{-1}}$ and $86\,\mathrm{km\,s^{-1}}$, respectively. The modal velocity ($\sim500\,\mathrm{km\,s^{-1}}$) of stripped material in the N100 model exceeds that of the W7 model ($\sim350\,\mathrm{km\,s^{-1}}$) by a factor of 1.4. We therefore conclude that our numerical results of ejecta-\otherstar{} interaction are not significantly dependent on the studied two near-Chandrasekhar-mass explosion models.

\section*{Acknowledgements}

We thank the anonymous referee for their comments that helped to improve this paper. CM thanks the University of Surrey and Yunnan Observatories for their guidance and support during his foreign-exchange year.  ZWL is supported by the National Natural Science Foundation of China (NSFC, Nos.~11873016, 11973080, and 11733008), the National Key R\&D Program of China (Nos. 2021YFA1600400, 2021YFA1600401), the Chinese Academy of Sciences, Yunnan Fundamental Research Projects (Grant No. 202001AW070007) and Yunnan Province (Nos.~12090040, 12090043, 2019HA012, and 2017HC018) and thanks the University of Surrey \emph{Institute of Advanced Studies} for a visiting fellowship.  RGI was supported by STFC grants ST/L003910/1 and ST/R000603/1, and thanks Yunnan Observatories for their gracious hospitality during exchange visits.   KCP is supported by the Ministry of Science and Technology of Taiwan through grants MOST 107-2112-M-007-032-MY3 and MOST 110-2112-M-007-019. HLC is supported by the National Natural Science Foundataion of China (NSFC, Nos.12073071) and Yunnan Fundamental Research  Projects  (Grant Nos.~202001AT070058, 202101AW070003), the science  research grants from the China Manned Space Project with No. CMS-CSST-2021-A10  and Youth Innovation Promotion Association of Chinese Academy of Sciences (Grant no. 2018076). The authors gratefully acknowledge the ``PHOENIX Supercomputing Platform'' jointly operated by the Binary Population Synthesis Group and the Stellar Astrophysics Group at Yunnan Observatories, CAS. This work made use of the Heidelberg Supernova Model Archive (HESMA,  \url{https://hesma.h-its.org}, \citealt{Kromer2017}).

\section*{DATA AVAILABILITY}
The data underlying this article will be shared on reasonable request to the corresponding author.

\bibliographystyle{mnras}
\bibliography{ref}

\end{document}